\documentclass[superscriptaddress,twocolumn,prl]{revtex4-1}

\usepackage{graphicx}
\usepackage{amsmath}
\usepackage{amssymb}
\usepackage{amsthm}

\usepackage[utf8]{inputenc}

\usepackage{hyperref}
\usepackage[all]{hypcap}
\hypersetup{bookmarksnumbered=true,
	colorlinks = true,    
	allcolors = blue,            
	breaklinks = true,   
	backref = true               
}

\begin{document}

\title{Does a pristine, unreconstructed SrTiO$_3$(001) surface exist?}
\author{Igor Sokolović}
\affiliation{Institute of Applied Physics, TU Wien, Wiedner Hauptstrasse 8-10/134, 1040 Vienna, Austria}

\author{Giada Franceschi}
\affiliation{Institute of Applied Physics, TU Wien, Wiedner Hauptstrasse 8-10/134, 1040 Vienna, Austria}

\author{Zhichang Wang}
\altaffiliation[Current address: ]{Department of Chemistry, College of Chemistry and Chemical Engineering, Xiamen University, Xiamen 361005, China.}
\affiliation{Institute of Applied Physics, TU Wien, Wiedner Hauptstrasse 8-10/134, 1040 Vienna, Austria}

\author{Jian Xu}
\altaffiliation[Current address: ]{College of Materials Science and Engineering, Chongqing University, Chongqing 400044, China.}
\affiliation{Institute of Applied Physics, TU Wien, Wiedner Hauptstrasse 8-10/134, 1040 Vienna, Austria}

\author{Jiří Pavelec}
\affiliation{Institute of Applied Physics, TU Wien, Wiedner Hauptstrasse 8-10/134, 1040 Vienna, Austria}

\author{Michele Riva}
\affiliation{Institute of Applied Physics, TU Wien, Wiedner Hauptstrasse 8-10/134, 1040 Vienna, Austria}

\author{Michael Schmid}
\affiliation{Institute of Applied Physics, TU Wien, Wiedner Hauptstrasse 8-10/134, 1040 Vienna, Austria}

\author{Ulrike Diebold}
\affiliation{Institute of Applied Physics, TU Wien, Wiedner Hauptstrasse 8-10/134, 1040 Vienna, Austria}

\author{Martin Setvín}
\email[Corresponding author: ]{setvin@iap.tuwien.ac.at}
\affiliation{Institute of Applied Physics, TU Wien, Wiedner Hauptstrasse 8-10/134, 1040 Vienna, Austria}
\affiliation{Department of Surface and Plasma Science, Faculty of Mathematics and Physics, Charles University, 180 00 Prague 8, Czech Republic}
\begin{abstract}

The surfaces of perovskite oxides affect their functional properties, and while a bulk-truncated (1$\times$1) termination is generally assumed, its existence and stability is controversial. Here, such a surface is created by cleaving the prototypical SrTiO$_3$(001) in ultra-high vacuum, and its response to thermal annealing is observed. 
Atomically resolved nc-AFM  shows that intrinsic point defects on the as-cleaved surface migrate at temperatures above 200\,$^\circ$C. At 400--500\,$^\circ$C, a disordered surface layer forms, albeit still with a (1$\times$1) pattern in LEED.  Purely TiO$_2$-terminated surfaces, prepared by wet-chemical treatment, are also disordered despite their (1$\times$1) periodicity in LEED.

\end{abstract}

\maketitle

\begin{figure*}[t]
	\begin{center}
		\includegraphics[width=2.0\columnwidth,clip=true]{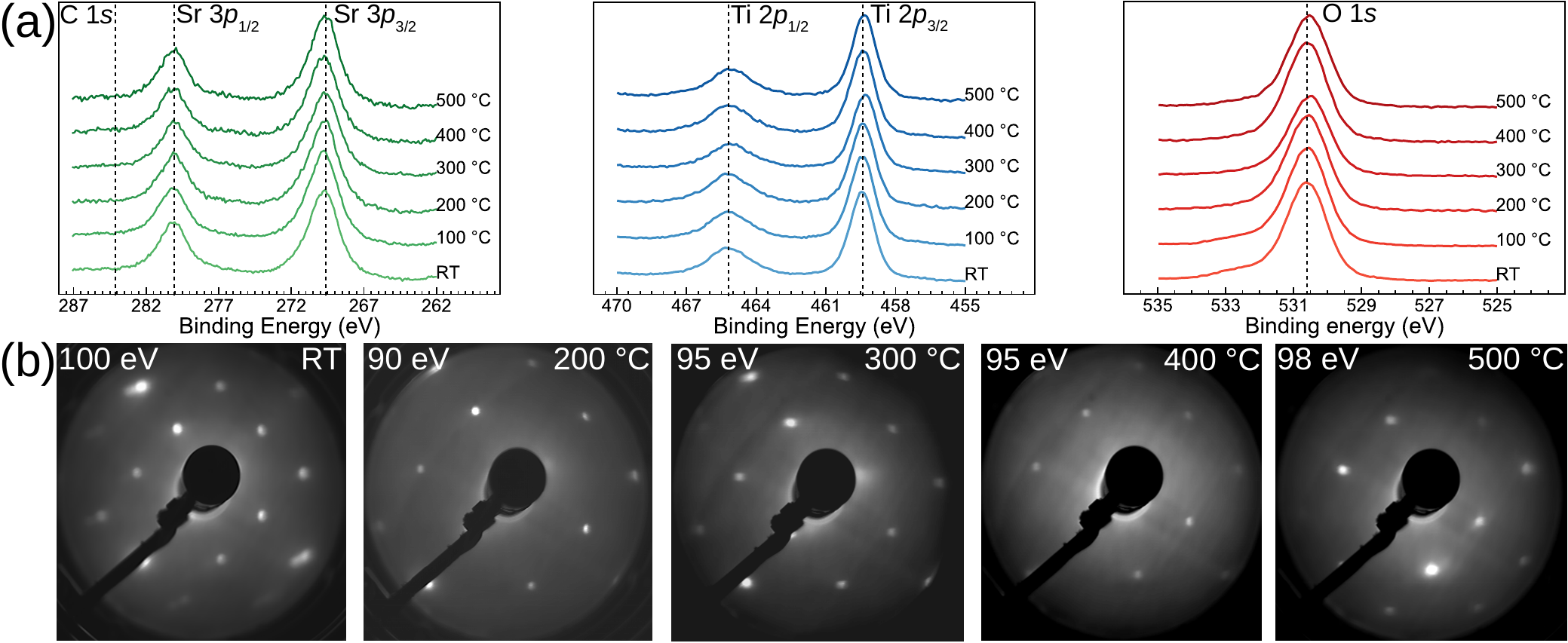}
	\end{center}
	\caption{			
		 SrTiO$_3$(001)-(1$\times$1), cleaved at room temperature and annealed to increasing temperatures.
		(a) XPS spectra of the main core levels, obtained with Mg $K\alpha$ radiation. Spectra are offset vertically for clarity.  (b) LEED patterns recorded at room temperature after cleavage, and after various annealing steps.  No substantial change is discerned in these area-averaging measurements. 
	}\label{fig1}
\end{figure*}

\begin{figure*}[t]
	\begin{center}
		\includegraphics[width=2.0\columnwidth,clip=true]{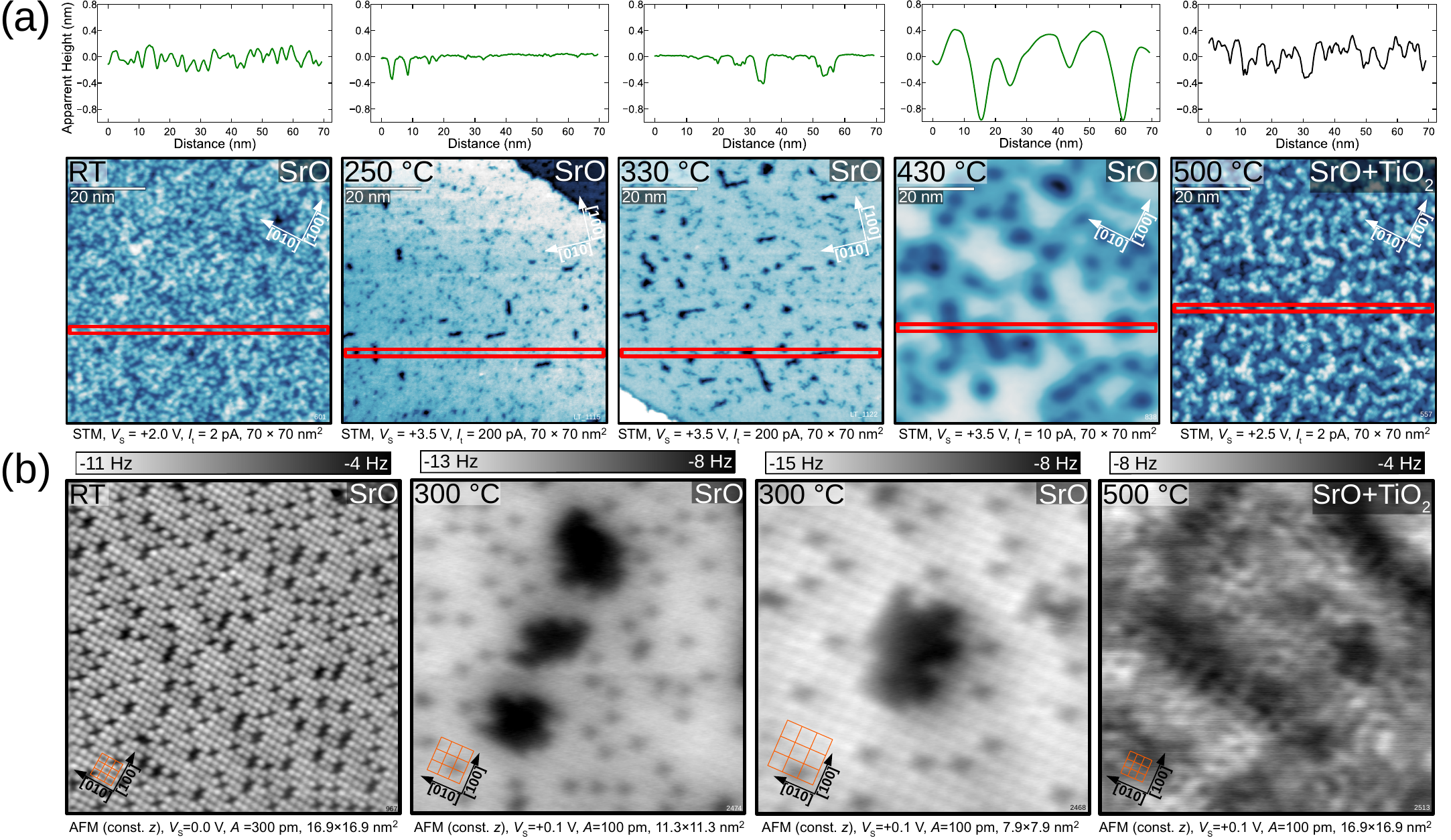}
	\end{center}
	\caption{
	SrO termination of the bulk-terminated SrTiO$_3$(001) surface, as-cleaved and annealed at different temperatures. Sequence of (a) empty-states STM images with line scans extracted from the marked red rectangles;  (b)  atomically resolved constant-height AFM images, with the (1$\times$1) grid shown in orange. 
}\label{fig2}
\end{figure*}

\begin{figure*}[t]
	\begin{center}
		\includegraphics[width=2.0\columnwidth,clip=true]{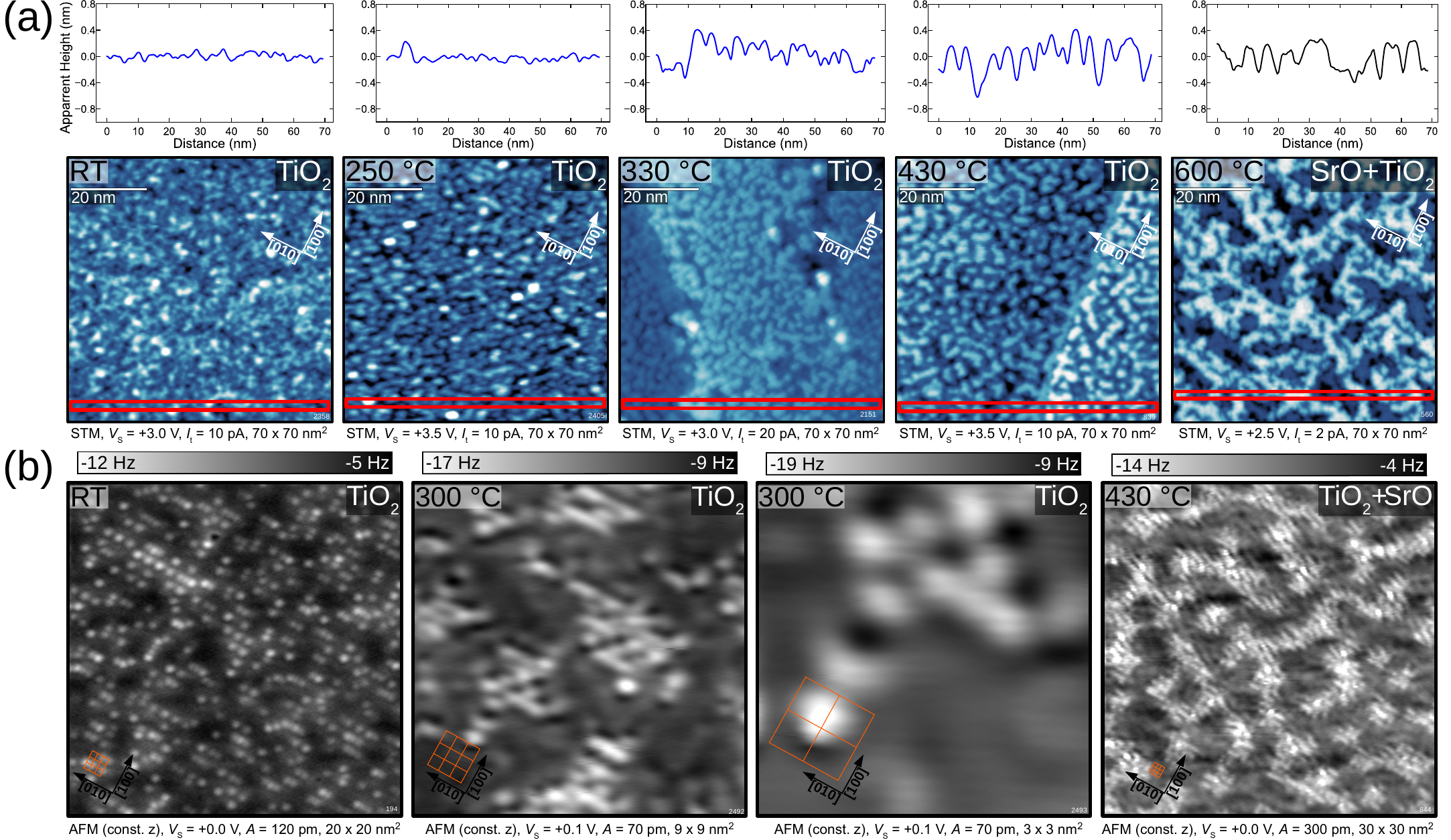}
	\end{center}
	\caption{
	TiO$_2$ termination of the bulk-terminated SrTiO$_3$(001) surface, as-cleaved and annealed at different temperatures. Sequence of (a) empty-states STM images and line scans extracted from the marked red rectangle;  (b) atomically resolved constant-height AFM images, with the (1$\times$1) grid shown in orange.
}\label{fig3}
\end{figure*}




Materials crystallizing in the perovskite structure attract attention in many fields due to their high tunability and the resulting, broad spectrum of physical and chemical properties \cite{hwang2017perovskites,zhu2014perovskite,ha2011adaptive}. Most applications of these materials rely on their surfaces and interfaces. The ternary chemical composition  of perovskites allows for many different surface terminations \cite{erdman2003surface, kienzle2011vacant} with dramatically different chemical and physical behaviors.  It is mostly assumed, however, that perovskites prepared by wet chemical techniques are bulk-terminated (1$\times$1) \cite{kawasaki1994atomic}. Theoretical modeling of catalytic processes \cite{Suntivich2011, Norskov2011} and physical phenomena \cite{santander-syro2011two, meevasana2011creation} generally assume this pristine surface structure.  The interfaces between two perovskites preserve the bulk crystal structure and the properties of the junction are strongly influenced by the local defect chemistry. Several emerging applications also profit from the bulk termination, such as employing ferroelectricity for optimizing or promoting catalytic reactions \cite{Kakekhani2015ACS, Kakekhani2016}.

Surprisingly, the available atomic-scale information on bulk-terminated cubic perovskites is quite limited. A prototypical example is SrTiO$_3$: The most common SrTiO$_3$(001) surface preparation method is wet-chemical treatment.  Etching with buffered-HF \cite{kawasaki1994atomic}  preferentially removes the SrO layer and leaves the surface fully covered with flat, TiO$_2$-terminated terraces \cite{koster1998quasi, koster1998influence}. This method was re-examined several times \cite{deak2006ordering,castell2002scanning,silly2006srtio3,baniecki2008photoemission,herger2007surfacePRL}; as was shown by x-ray photoelectron spectroscopy (XPS)  it can induce an unintentional substitution of oxygen with F \cite{chambers2012unintentional}. This can be avoided by etching the surface with HCl--HNO$_3$ \cite{kareev2008atomic} or non-acidic solvents  \cite{chambers2012unintentional,connell2012preparation, gerhold2014stoichiometry}. Etched surfaces must undergo annealing to at least 600\,$^\circ$C after introduction into vacuum to remove adsorbates. Up to this temperature, perfectly flat, TiO$_2$-terminated surfaces usually display a (1$\times$1) pattern in low-energy electron diffraction (LEED) or reflection high energy diffraction (RHEED) \cite{castell2002scanning,di2012observation}, while a series of surface reconstructions appear after annealing at higher temperatures \cite{herger2007surfacePRL,herger2007surface,silly2006srtio3,dagdeviren2016surface}. Some  reconstructions have been atomically resolved using scanning tunneling microscopy (STM) including the (1$\times$2), (2$\times$2), \textit{c}(4$\times$2), \textit{c}(4$\times$4), and \textit{c}(6$\times$2) \cite{castell2002scanning,erdman2003surface,lanier2007atomic,gerhold2014stoichiometry}, and several more on  surfaces that were sputtered and annealed in ultrahigh vacuum (UHV) \cite{deak2006ordering}.

So far, the structural characterization of SrTiO$_3$(001) and other cubic perovskites was mostly based on diffraction techniques. A simple (1$\times$1) diffraction pattern could also stem from the bulk underneath a disordered layer, however; a true proof of a crystalline top surface requires atomically resolved imaging.  For perovskite oxides this has  been demonstrated only recently \cite{sokolovic2019incipient,setvin2018polarity} using non-contact atomic force microscopy (nc-AFM, AFM hereafter) \cite{morita2015noncontact}. This technique  provides clear, atomically resolved images of the (1$\times$1) termination, whereas STM shows no atomic corrugation \cite{sokolovic2019incipient}. It should be noted that the cleaving process that produces such unequivocally crystalline (1$\times$1) terminated SrTiO$_3$(001) surfaces relies on incipient ferroelectricity, and that the induced polarity necessarily results in a sizable density of charged point defects \cite{sokolovic2019incipient}.  Such cleaved surfaces can have micron-sized domains with exclusively TiO$_2$ and SrO termination, and each contains 14$\pm$2\% Sr adatoms and Sr vacancies, respectively.

One could expect that thermal annealing heals such intrinsic point defects. Instead, this work shows that the as-cleaved SrTiO$_3$(001)-(1$\times$1) surface is unstable. Raising the temperature results in the lateral migration of the point defects, and an overlayer without long-range order forms above 400\,$^\circ$C.  Having established that AFM is capable of providing a clear picture of ordered perovskite surfaces, it was applied to SrTiO$_3$(001) prepared by wet etching.  These samples show no signs of an ordered surface in AFM, however, raising doubts that the commonly observed (1$\times$1) diffraction pattern indicates a crystalline, unreconstructed surface. 



\begin{figure}[h!]
	\begin{center}
		\includegraphics[width=1.0\columnwidth,clip=true]{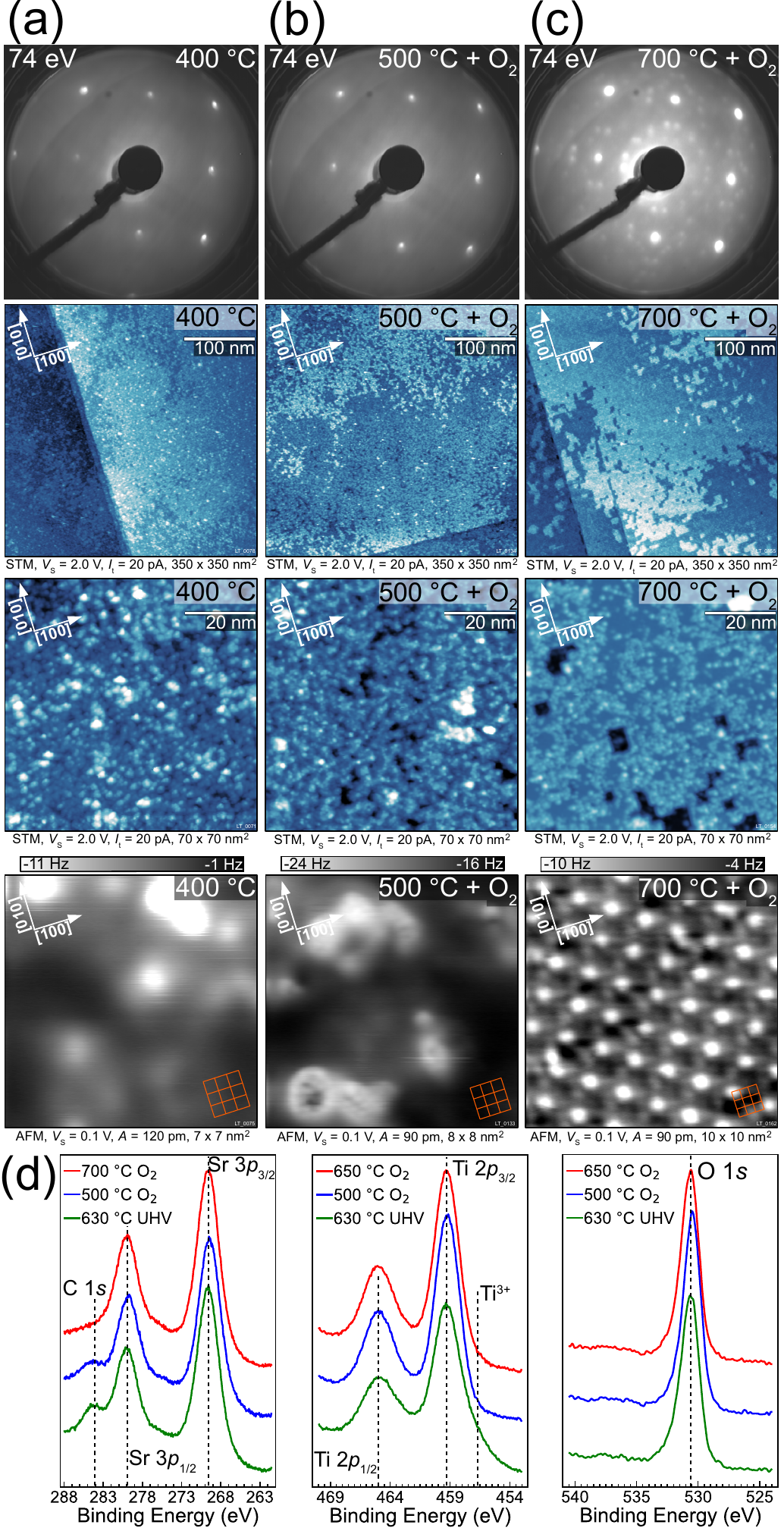}
	\end{center}
	\caption{
		Cut-and-polished TiO$_2$-terminated SrTiO$_3$(001) surface prepared by  \textit{ex-situ} wet-chemical treatment. (Top to bottom) LEED, large- and small-area empty-states STM, and AFM images after annealing at (a) 400\,$^\circ$C in UHV, (b) 500\,$^\circ$C and (c) 700\,$^\circ$C in 1$\times$10$^{-6}$\,mbar O$_2$. (d) XPS core-level spectra of the main elements at different annealing conditions, obtained with Al$K\alpha$ line in grazing emission.
	}\label{fig4}
\end{figure}




SrTiO$_3$ samples doped with 0.7\,at.\% Nb were used. For cleaving, samples were held in a cleaving device \cite{sokolovic2019incipient} constructed from Mo. After insertion into a UHV chamber with a base pressure below 2$\times$10$^{-10}$ mbar, the device was thoroughly degassed and cooled down to room temperature (RT) prior to cleaving. Annealing was performed \textit{via} a resistive-heating wire below the sample mount. The precision of the reported annealing temperatures is estimated as $\pm$30\,$^\circ$C due to the limited thermal conductivity of the sample mount. Annealing times ranged from 30 to 60 min. 

Scanning--probe measurements were performed in UHV with a base pressure below 2$\times$10$^{-11}$~mbar on a ScientaOmicron low-temperature STM/AFM at either $T$=77.7\,K or $T$=4.8\,K, using  qPlus cantilevers \cite{giessibl2013sensor} with a separate wire for the tunneling current \cite{Majzik2012} and a differential preamplifier \cite{GiessiblPreamp}. Etched tungsten tips were glued on the tuning fork and cleaned by self-sputtering in Ar atmosphere \cite{setvin2012ultrasharp} prior to the experiment.  The resonance frequency of the  qPlus cantilevers ranged from 25 to 80\,kHz, with Q factors of $\approx$ 50000. The STM images were obtained by applying a positive bias to the sample. All presented AFM images were acquired with an oxygen-terminated tip \cite{SokolovicPNAS2020}, $i.e.$, surface cations appear attractive (bright) and anions repulsive (dark) \cite{Yurtsever2012}. 

Figure\,\hyperref[fig1]{1} shows XPS and LEED  results of SrTiO$_3$(001) prepared by the \emph{in-situ} cleaving \cite{sokolovic2019incipient} and after annealing at increasingly higher temperatures  for 45 min each, up to 500\,$^\circ$C. In XPS the surfaces are free of contaminants such as carbon.  The core-level spectra of the constituents do not change, indicating that the overall surface stoichiometry is preserved.  Despite the reducing character of the UHV environment, the elements retain their oxidation state. 

The as-cleaved surface exhibits a clear (1$\times$1) LEED pattern. The periodicity does not change various annealing steps. On the other hand, the sharpness of the LEED spots and the background intensity vary: The sharpest diffraction pattern was observed after annealing at 200\,$^\circ$C, while further annealing degraded the pattern somewhat. A distinct (1$\times$1) LEED pattern is also reported throughout literature for polished SrTiO$_3$(001) surfaces annealed at temperatures above 600\,$^\circ$C, necessary for degassing after introduction into UHV. Above 850\,$^\circ$C, spots originating from \textit{c}(4$\times$2)  \cite{erdman2002structure,erdman2003srtio3}, or (1$\times$2) \cite{castell2002scanning} reconstructions start to appear.

The surfaces of cleaved samples have SrO  and TiO$_2$ domains that span 10 to 100 $\mu$m \cite{sokolovic2019incipient}. This is below the resolution of either of the techniques applied in Fig.\,\hyperref[fig1]{1}.  The STM/AFM results in Figs. \hyperref[fig2]{2} and \hyperref[fig3]{3} show the temperature evolution of the two terminations separately. 




Figure\,\hyperref[fig2]{2} focuses on the SrO termination. STM images illustrate the large-scale surface morphology, while smaller-size AFM images provide details of the atomic structure. After cleaving at room temperature the surface is atomically flat \cite{sokolovic2019incipient}. The SrO (1$\times$1) surface is covered with the specific concentration of 14$\pm$2\% of Sr vacancies, V$_{\mathrm{Sr}}^{2-}$, apparent as black, `missing' atoms in Fig.\,\hyperref[fig2]{2(b)} \cite{sokolovic2019incipient}. The corresponding STM image shows a corrugation as high as a full unit cell ($\approx$0.4\,nm), which is a purely electronic effect originating from the band bending induced by the charged Sr vacancies.  

Annealing at 250$^\circ$C results in the formation of pits at the flat SrO terraces, seen in both,  STM and AFM. Their surroundings remain unreconstructed, with fewer V$_{\mathrm{Sr}}^{2-}$ defects. Intrinsic V$_{\mathrm{Sr}}^{2-}$ agglomerate into larger  half-unit-cell-deep pits, that expose the underlying TiO$_2$ termination. No oxygen atoms are visible within the pits; they can be considered aggregates of Schottky-type defects \cite{walsh2015self}, formed by Sr vacancy diffusion and the formation of O vacancies \cite{walsh2011strontium}.  STM images of the SrO domains annealed to 250\,$^\circ$C and 330\,$^\circ$C show significantly smaller corrugation compared to the as-cleaved surface, consistent with a reduced band bending. Such surfaces also show a LEED pattern [Fig.\,\hyperref[fig1]{1(b)}] with the lowest background intensity.

Further annealing results in the lateral growth of the pits, preferentially along the [100] and [010] directions. The STM images in  Fig.\,\hyperref[fig2]{2(a)} appear considerably rougher (up to 3 layers). Annealing the SrO termination at 500\,$^\circ$C results in a loss of the (1$\times$1) ordering in AFM. Instead, images locally show a short-range (2$\times$2) periodicity, while LEED  retains the (1$\times$1) symmetry.  The surface corrugation deduced from the AFM images is less than half unit cell. STM images exhibit larger apparent height differences, but this can be partially attributed to electronic effects, when domains with different electronic properties form. At this stage of annealing, it is no longer possible to distinguish the previously SrO- and TiO$_2$-terminated areas: The entire cleaved SrTiO$_3$(001) surface shows the same morphology in STM and AFM images as in the rightmost panels of Fig.\,\hyperref[fig2]{2a} and Fig.\,\hyperref[fig2]{2b}, respectively. The (1$\times$1) LEED pattern of this surface is attributed to diffraction from the subsurface layers, while the disordered surface layer results in the increased background intensity.




The evolution of the TiO$_2$ termination with temperature is shown in Fig.\,\hyperref[fig3]{3}. After cleaving, the TiO$_2$ termination hosts Sr$_{\mathrm{ad}}^{2+}$ adatoms [bright dots in Fig.\,\hyperref[fig3]{3(b)}], complementary to the V$_{\mathrm{Sr}}^{2-}$ vacancies at the SrO termination. In STM, the two terminations appear very similar. Annealing above 250\,$^\circ$C results in the formation of small, disconnected islands. The clustering of Sr$_{\mathrm{ad}}^{2+}$ adatoms requires the presence of O$^{2-}$  to compensate their electric charge, and indeed AFM images show the presence of anions [dark dots in Fig.\,\hyperref[fig3]{3(b)}]. These SrO islands show tiny areas with a \textit{c}(2$\times$2) SrO structure on top of the TiO$_2$ termination. The intrinsic excess of Sr adatoms at the as-cleaved TiO$_2$ termination constitute an ideal seed for the crystal growth of the next perovskite SrTiO$_3$ layer, provided the temperature is sufficiently high for the diffusion of the adatoms and of oxygen to complete the SrO stoichiometry.

The \textit{c}(2$\times$2)-like areas spread with increasing temperature up to 430\,$^\circ$C. Islands grow more connected, while still not covering the entire surface. The maximum coverage of this SrO superstructure over the TiO$_2$ termination is limited by the initial 0.14~ML coverage of the Sr adatoms. When arranged in a  \textit{c}(2$\times$2) superstructure, it can cover 28\% of the surface area, close to the maximum coverage observed in AFM. After annealing at 500\,$^\circ$C the previous TiO$_2$ termination appears disordered and becomes indistinguishable from what was the SrO termination [rightmost panel of Fig.\,\hyperref[fig2]{2(a)} and Fig.\,\hyperref[fig2]{2(b)}].  Further annealing of the mixed-termination morphology at 600\,$^\circ$C [rightmost panel of Fig.\,\hyperref[fig3]{3(a)}] does not improve the surface roughness, but instead increases the width of the pits and islands.

The two opposite terminations of the as-cleaved SrTiO$_3$(001)-(1$\times$1) surface experience a complementary evolution with annealing. The pit/island creation mechanism is induced by the presence of the intrinsic, polarity--compensating point defects, $i.e.$, Sr vacancies and adatoms \cite{sokolovic2019incipient}. Migration of these charged  V$_{\mathrm{Sr}}^{2-}$ and Sr$_{\mathrm{ad}}^{2+}$ point defects is activated at temperatures as low as 200\,$^\circ$C. Moreover, when the surface is additionally enriched by Sr adatoms \textit{via} evaporation, the adatoms are mobile at both terminations and start to aggregate at temperatures as low as 150\,$^{\circ}$C, as shown in the Supplemental Material (SM). The disappearance/appearance of O likely originates from the exchange with the subsurface region, because lateral diffusion across the whole micrometers-wide domains is unlikely \cite{riva2019pushing,riva2019epitaxial}. The temperature range for migration of vacancies is slightly lower than reported in the literature for SrTiO$_3$ \cite{de2015oxygen, riva2018influence}, but can be rationalized by the presence of electric fields related to the charged defects.



Since the intrinsic point defects dominate the thermal behavior of the cleaved crystals, annealing excursions were also conducted on SrTiO$_3$ samples prepared by a wet chemical treatment that exposes only the TiO$_2$ termination.  Full details are laid out in the SM. Two cut-and-polished SrTiO$_3$(001) crystals, again with 0.7\,at.\% Nb doping, were cleaned \textit{ex situ} and boiled in ultra-pure water to etch away the soluble SrO termination. One sample was baked in air at 950\,$^\circ$C prior to introduction to UHV (to create large, flat terraces) and turned out to be contaminated with carbon, alkali, and alkaline earth metals, see SM. The second sample was introduced to UHV directly after wet cleaning and was contaminated with carbon alone.

Figure\,\hyperref[fig4]{4} shows the temperature evolution of the latter sample.  After annealing to mild temperatures, LEED shows a distinct (1$\times$1) pattern, and large-area STM images show flat terraces.   In AFM, however, clumps are visible, likely due to contamination.  A substantial C1s signal in XPS [Fig.\,\hyperref[fig4]{4(d)}], indicates that the contaminants are carbon-based organics.  Annealing at 500\,$^\circ$C--650\,$^\circ$C in 1$\times$10$^{-7}$--1$\times$10$^{-6}$ O$_2$ back-pressure for 1--2 h gradually removed the C, but  constant-height AFM still shows a surface covered by undetermined hillocks, with no hints of the underlying substrate.  A significant reduction of C occurred only after annealing at 700\,$^\circ$C in 1$\times$10$^{-7}$\,mbar O$_2$ for 2 h. This treatment resulted in a reconstructed surface with a $\left(\sqrt{\mathrm{13}}\times \sqrt{\mathrm{13}}\right)\!\!R\mathrm{33.7}^\circ$ superstructure \cite{kienzle2011vacant,ohsawa2016negligible} clearly visible in  LEED  and the constant-height AFM image in Fig.\,\hyperref[fig4]{4(c)}.




In summary, the search for an SrTiO$_3$(001)-(1$\times$1) surface that can be considered 'pristine' --- crystalline and well-ordered, with a negligible amount of defects and contaminants --- was not successful so far.  As-cleaved surfaces come closest, but necessarily contain both, SrO- and TiO$_2$-terminated domains (albeit of considerable size), and charged point defects.  The temperature-induced transformation to an ill-defined, disordered top layer, indicates that the (1$\times$1) has a high surface energy and is only metastable. The techniques generally applied to judge surface quality (electron diffraction, XPS, large-area STM or ambient AFM) give results that would be consistent with a perfect (1$\times$1) termination; but are contradicted by atomically resolved nc-AFM.  The same is true for the etched TiO$_2$-terminated surfaces that are used to great extent as substrates in the growth of heteroepitaxial oxide films.  While such surfaces can display a flat morphology with a high-quality diffraction pattern, nc-AFM shows no signs of the unreconstructed surface. The temperature required for removing carbon lies above the stability region of the bulk-terminated surface.

The (001) surface  of SrTiO$_3$ and other perovskites continue to be of great interest for both, probing their intrinsic properties and as substrates for heteroepitaxy.  As the results presented here show, the assumption of an atomically clean, crystalline bulk-termination might not be warranted. This should be considered in the proper interpretation of experimental data.





This work was supported by the Austrian Science Fund (FWF) projects Wittgenstein Prize (Z-250), Solids4Fun (F-1234) and SuPer (P32148-N36). J.X. acknowledges the support from the National Natural Science Foundation of China (91634106), China Scholarship Council and Chongqing University. M.Se. acknowledges support from the Czech Science Foundation GACR 20-21727X and GAUK Primus/20/SCI/009.


\bibliography{Bibliography}
\pagebreak{}



\end{document}